\documentclass[USenglish,oneside,twocolumn]{article}

\usepackage[utf8]{inputenc}
\usepackage[big]{dgruyter_NEW}
 
\DOI{foobar}

\cclogo{\includegraphics{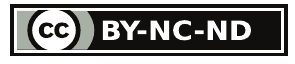}}

\author*[1]{Maaz Bin Musa}
\author[2]{Rishab Nithyanand}

\affil[1]{University of Iowa, Email: \texttt{maazbin-musa@uiowa.edu}, Web: \url{https://maazbinmusa.github.io/}}
\affil[2]{University of Iowa, Email: \texttt{rishab-nithyanand@uiowa.edu}, Web: \url{https://sparta.cs.uiowa.edu/people/rishabn}}

\usepackage{todonotes}
\usepackage{marginnote}

\usepackage{blindtext}
\usepackage{graphicx,balance,multirow,xspace,subcaption,longtable}
\usepackage{makecell,amsmath,amssymb,cleveref,algorithm}
\usepackage[font={small}, textfont=md]{caption}
\usepackage[T1]{fontenc}
\usepackage[draft,inline,nomargin,index]{fixme}
\usepackage{booktabs}
\usepackage{enumitem}
\usepackage[noend]{algpseudocode}

\PassOptionsToPackage{hyphens}{url}
\usepackage{hyperref}
\newcolumntype{P}[1]{>{\arraybackslash}p{#1}}
\newcolumntype{X}[1]{>{\centering\arraybackslash}p{#1}}

\fxsetup{theme=colorsig,mode=multiuser,inlineface=\itshape,envface=\itshape}

\usepackage{pifont}

\newcommand\clearrow{\global\let\rowmac\relax}

\newcommand{\para}[1]{{\vspace{.05in} \bf \noindent #1 }}
\newcommand{\parait}[1]{{\vspace{.05in} \it \noindent #1 }}

\crefformat{section}{\S#2#1#3}
\crefformat{subsection}{\S#2#1#3}
\crefformat{subsubsection}{\S#2#1#3}

\renewcommand{\footnotesize}{\scriptsize}
\newcommand{\etc}{etc.}
\newcommand{\eg}{e.g.,\ }
\newcommand{\etal}{et al.\xspace}
\newcommand{\ie}{i.e.,\ }

\newcommand{\atomarchive}{\url{https://vitalstatistix.cs.uiowa.edu:2443/maaz/atom-archive}}

\definecolor{applegreen}{rgb}{0.55, 0.71, 0.0}

\begin{document}

\title{\huge ATOM: Ad-network Tomography}
\runningtitle{A Generalizable Technique for Inferring Tracker-Advertiser Data
Sharing in the Online Behavioral Advertising Ecosystem}
\begin{abstract}
{
  Data sharing between online trackers and advertisers is a key component in
  online behavioral advertising. This sharing can be facilitated through
  a variety of processes, including those not observable to the user's browser.
  The unobservability of these processes limits the ability of researchers and
  auditors seeking to verify compliance with recent regulations (\eg CCPA and
  CDPA) which require complete disclosure of data sharing partners.
  Unfortunately, the applicability of existing techniques to make inferences
  about unobservable data sharing relationships is limited due to their
  dependence on protocol- or case-specific artifacts of the online behavioral
  advertising ecosystem (\eg they work only when client-side header bidding is
  used for ad delivery or when advertisers perform ad retargeting).
  As behavioral advertising technologies continue to evolve rapidly, the
  availability of these artifacts and the effectiveness of transparency
  solutions dependent on them remain ephemeral.
  \\
  In this paper, we propose a generalizable technique, called ATOM, to infer
  data sharing relationships between online trackers and advertisers. ATOM is
  different from prior approaches in that it is universally applicable ---
  \ie independent of ad delivery protocols or availability of artifacts.
  ATOM leverages the insight that by the very nature of behavioral advertising,
  ad creatives themselves can be used to infer data sharing between trackers
  and advertisers --- after all, the topics and brands showcased in an ad are
  dependent on the data available to the advertiser.
  Therefore, by selectively blocking trackers and monitoring changes in the
  characteristics of ad creatives delivered by advertisers, ATOM is able to
  identify data sharing relationships between trackers and advertisers.
  The relationships discovered by our implementation of ATOM include those not
  found using prior approaches and are validated by external sources.
}
\end{abstract}


\journalname{Proceedings on Privacy Enhancing Technologies}
\DOI{Editor to enter DOI}
\startpage{1}
\received{..}
\revised{..}
\accepted{..}

\journalyear{..}
\journalvolume{..}
\journalissue{..}

\maketitle
\sloppy

\section{Introduction}\label{sec:introduction}

\para{Investment in online behavioral advertising is growing rapidly.}
%
%
Over the past decade, the advertising industry has demonstrated a clear
preference for online behavioral advertising --- \ie displaying individually
targeted ads based on what is known about the users' online habits and
behaviors. In fact, recent reports by eMarketer \cite{eMarketer-2022} and the
Interactive Advertising Bureau (IAB) \cite{IAB-2021} have estimated the 2021
programmatic digital ad spend in the United States alone to be between
\$160-211B and in excess of \$315B by 2025. This represents a 
10-18\% year-over-year growth and over two-thirds of the ad spend across all media
(including television, print, and radio) \cite{eMarketer-2019}. 

\para{Online behavioral advertising has led to the commodification of data.}
%
%
In online behavioral advertising, brands place bids for ad slots made available
by online publishers when a user loads their service. The logistics of the
associated bidding process are often fully outsourced to programmatic
advertising organizations called Demand Side Platforms (DSPs or {\em
advertisers}). These advertisers allow bids to be made and their associated
dollar values to be computed in real time. In simple terms, bid values are
dependent on the ad slot (\eg the location and size of the slot) and the
likelihood of the attracting the engagement of the specific user (estimated by
known user characteristics such as their interests, location, purchase habits,
\etc). Bearing in mind our simplified overview ignores many complexities of 
the advertising ecosystem.

%
%
For brands and advertisers, such targeting has been shown to be significantly
more cost-effective than traditional (\ie contextual) advertising
\cite{Jivox-2016} when presented with high quality user data. Today, the
online advertising landscape contains an entire data ecosystem focused on
harvesting and trading user data. This online data ecosystem satisfies
advertisers' need for high quality user data and publishers' dependence on
advertising revenue.
%
%
Unfortunately, this commodification of user data has resulted in the
development of privacy-invasive user tracking practices (\eg stateless tracking
methods such as browser fingerprinting \cite{Laperdrix-TWEB, Iqbal-SP2021}) and
the emergence of data sharing relationships between online entities that are
unknown to the user (\eg through cookie syncing \cite{Acar-CCS2014,
Papadopoulos-WWW2019}, server-side sharing \cite{Tune-2014, Affise-2022,
UnifiedID}, or third-party data brokers \cite{DataBrokers-FastCompany2019,
DataBrokers-PRCH2022}).

\para{Emerging regulations around data sharing will be hard to enforce.}
%
%
In recent years, regulators have taken notice of the increasing concerns
surrounding the lack of online privacy controls \cite{Privacy-FTC1998,
Privacy-Pew2019}, the opacity of online user data harvesting and sharing
\cite{Brokers-FTC2014}, and the failure of the advertising ecosystem's
self-regulation efforts \cite{Hoofnagle-CP2005,Dixon-WPF2007,Ginosar-PI2014}.
This has resulted in the passage of numerous privacy-focused regulations that
explicitly limit how user data  might be gathered, handled, and shared. Most
relevant to our work are the EU's GDPR \cite{GDPR}, California's CCPA
\cite{CCPA} and CPRA \cite{CPRA}, and Virginia's CDPA \cite{CDPA}. Each of
these regulations attempt to improve transparency in the online data ecosystem
by requiring organizations to disclose the sale or sharing of non-public
consumer data to third-parties. 
%
%
Unfortunately, despite a few successes in reining in the online data
ecosystem's data gathering practices \cite{GoogleFine-CNIL2022,
FacebookFine-CNIL2022}, these regulations are hard to enforce for two key
reasons. First, the absence of private right of action
\cite{Goldman-SCULSRP2020} creates bottlenecks at the enforcement agencies.
Second, many aspects of the regulations are not amenable to large-scale
auditing systems that can verify regulatory compliance
\cite{Razaghpanah-NDSS2018}. For example, the enforcement of regulations
surrounding data sharing will be largely ineffective due to the absence of
general techniques to identify evidence of incorrectness or incompleteness of
disclosed data sharing relationships. 

\para{Identifying data sharing relationships is difficult.}
%
%
Broadly, data sharing in the online advertising ecosystem occurs between online
trackers (who gather user data) and online advertisers (who obtain this data
for the purposes of targeted advertising). This sharing can be facilitated
either by real time direct communication between the two entities (\ie
server-side data sharing), by real time re-directed communication facilitated
through the user's browser (\ie client-side sharing), or indirectly through
data brokers and other middlemen (\ie indirect sharing).  We describe
each of these data sharing mechanisms in more detail in \Cref{sec:background}.
%
%
The main challenge facing researchers seeking to measure these relationships is
the absence of a suitable vantage point from which to observe or infer sharing.
More specifically, measurements of the advertising ecosystem are typically only
afforded a view of interactions passing through the user's browser.
Consequently, only client-side sharing can be observed and recorded.
Unfortunately, current data sharing mechanisms are increasingly migrating
towards server-side sharing or indirect sharing due to recent browser policies
that block third-party cookies. Therefore, the relationships facilitated
through server-side remain opaque to researchers and
auditors.

\para{The applicability of current approaches to measure server-side and
indirect data sharing are limited.}
%
%
Interactions that facilitate data sharing between trackers and advertisers that
do not involve the user's browser are generally invisible to researchers and
auditors. Therefore, any mechanism to infer unobservable data sharing must
exclusively rely on client-observable side channels. Current approaches have
leveraged the side channels associated with specific artifacts of advertising
and ad delivery protocols. Unfortunately, these artifacts are not widespread
and limit the capabilities of the approaches that depend on them. 
%
%
For example, Cook \etal \cite{Cook-PETS2020} used client-observable advertiser
bids as a side channel through which data sharing relationships could be
inferred. However, advertiser bids are only visible when publishers
enable ad delivery through client-side Header Bidding --- an increasingly
uncommon situation given the ad industry's recent move towards server-side
Header Bidding \cite{HB-PubMatic2018, HB-Revamp2021, HB-AdExchanger2017}.
Working towards the same goal, Bashir \etal \cite{Bashir-Security2016} used
retargeted ads (\ie ads which showcase a product previously visited by a user)
to infer when server-side data sharing must have occurred. Once again, the
scope of this approach is limited to the identification of data sharing
relationships for retargeting. We highlight these prior approaches in detail in
\Cref{sec:related}.
%
%
{\em What is lacking is a generalizable (\ie artifact- and
mechanism-independent) approach to infer data sharing relationships between
trackers and advertisers.}

\para{We develop a generalizable technique to infer data sharing.}
%
%
At a high-level, our work is based on the insight that: {\em by their very
nature, personalized ads present an always-observable side channel that can be
used to infer data sharing relationships between advertisers and trackers}.
%
%
We arrive at this insight as follows:
\begin{itemize}
  \item
    Data sharing in the online advertising ecosystem occurs primarily
    to facilitate online behavioral advertising --- \ie to present personalized
    ads. Therefore, by the very nature of online behavioral advertising, the
    characteristics of the ad creatives (\eg topic, brands, \etc) displayed to
    a user must {\em always} be influenced by the data available to an
    advertiser. This presents an always-observable side channel.

  \item 
    As a consequence of the data-dependence of ad creatives, the creatives 
    presented by an advertiser having no data about a user will be
    significantly different than the creatives presented by an advertiser
    having rich data about a user's interests and browsing habits. For example,
    an advertiser that is aware of a user's interests in soccer will present
    different creatives and brands to the user than an advertiser with no
    knowledge of their interests.
  
  \item 
    The flow of information about a user's interests originates at online
    tracking organizations that harvest information about a user's online
    activities. This means that by blocking a tracking organization's ability
    to observe a user we also disrupt all information flows between that
    tracking organization and it's data sharing partners. Therefore, when the
    creatives presented by an advertiser are found to be statistically
    dependent on (un)blocking of a tracking organization, we have evidence of
    a data sharing relationship between the tracker and advertiser.
    
\end{itemize}

In this paper, we validate and operationalize this insight by testing the
following hypotheses.

\parait{H1. Characteristics of ad creatives are dependent on user interests
(\Cref{sec:interests}).} 
To test this hypothesis, we first create a number of online user personas
associated with specific interest groups (\eg sports, arts, \etc). 
Next, we use computer vision to extract the characteristics of the ad creatives
delivered to each of these personas when they visit a predefined set of
websites. 
Finally, we conduct statistical testing to identify if these extracted
characteristics are dependent on the persona interest group they were derived
from.
A positive finding of dependence would demonstrate that browsing history does
impact the extracted characteristics of delivered ad creatives --- thus
satisfying a necessary condition for using ad creatives to infer data sharing
relationships. 

\parait{H2. Characteristics of ad creatives can be used to infer
tracker-advertiser data sharing relationships (\Cref{sec:relationships}).}
We test this hypothesis by creating online user personas and gathering ad
creatives while systematically blocking popular tracking organizations from
observing the persona.
Next, we use statistical modeling to quantify the impact of blocking each
of the ten largest tracking organizations on the creatives delivered by each
advertiser.
We then use this measured impact to identify trackers that, when present,
significantly impact the delivered creatives of each advertiser. We infer the
presence of a data sharing relationship between these trackers and the
corresponding advertiser. 
Finally, we validate the correctness of our inferences using known
client-side sharing relationships and public disclosures made available through
CCPA's data broker registry \cite{CCPA-Brokers}.
\section{Background: Mechanisms of online behavioral advertising}
\label{sec:background}

Online behavioral advertising relies on two key processes working together in
unison: ad bidding and user data harvesting/dissemination. In this section, we
provide a high-level overview of these processes.

\subsection{Ad bidding mechanisms} \label{sec:background:bidding}
Online advertising seeks to provide users with ads tailored to their individual
interests. This requires coordination between publishers (who have the
attention of users) and brands (who seek to capture the attention of
`profitable' users) to occur in real time. Two programmatic mechanisms help
achieve this coordination: real-time bidding and header bidding.

\para{Real-Time Bidding (RTB).} RTB is currently the most popular mechanism
for trading ad inventory in real time \cite{RTB-ICO2019} and has been heavily
promoted by the Interactive Advertising Bureau since 2010 \cite{OpenRTB}. In
RTB, entities participate in the {\em supply side} or {\em demand side}.

\parait{The supply side.} When a user visits the website of a publisher, the
browser is directed to contact the publisher's ad-server to facilitate the
process of obtaining ads to display to the user. The publisher's request is
then forwarded, via a browser re-direct, to a {\em supply-side platform} (SSP)
along with information known to the publisher, as a first-party, about the user
(\eg user demographics, location, \etc) and the {\em floor price} for the
impression. The floor price represents the minimum value that the advertiser is
willing to accept for the ad slot. The SSP then augments this information
with it's own data about the user. This is possible since the browser redirect
includes the SSPs own cookie. This allows the SSP to create a more complete
picture of the advertising opportunity ({\em impression}). A summary of this
impression opportunity is then sent to an {\em ad exchange}.

\parait{The demand side.} Ad exchanges facilitate bidding on impression
opportunities. When an SSP shares an impression opportunity summary with an ad
exchange, the summary is shared with all clients of the ad exchange and the
bidding process is started. Clients of the ad exchange are typically {\em
demand-side platforms} (DSPs). These DSPs make programmatic bids on impression
opportunities on behalf of brands. In order to select an impression and a bid
value for their clients, DSPs leverage data obtained from the supply side,
third-party data brokers, and their own history with the user (made possible
through {\em cookie syncing} described in \Cref{sec:background:data}). If the
winning bid exceeds the floor price set by
the publisher, the ad exchange forwards the ad creative supplied by the winning
DSP to the SSP, charges the DSP a commission for the successful bid, and pays
the bid value to the SSP. The SSP forwards the ad creative to the publisher,
charges the publisher a commission for the successful impression, and pays the
bid value to the publisher.

\parait{When the publisher's floor price is not met.} In the event that the
publisher's SSP is unable to provide it with a bid that exceeds the floor price
set by the publisher, the publisher must repeat the entire RTB bidding process
with their next preferred SSP. This is repeated until the impression is sold to
a DSP. There are two major consequences of this `waterfall' approach used in
RTB: (1) publishers may not be able to sell their impression inventory without
extensive delays, (2) the impression may not be sold for it's true market value
(since not all bidders get to bid simultaneously).

\para{Header Bidding (HB).} Header Bidding is an emerging alternative to RTB.
Promoted by AppNexus, HB aims to provide more value for publishers and
advertisers by: (1) removing middlemen (\ie SSPs and ad exchanges) from the
bidding process and (2) flattening the waterfall approach of RTB
\cite{HB-AppNexus2017}. HB achieves this by: (1) allowing direct relationships
between publishers and any entity interested in the inventory of the publisher
and (2) soliciting bids from all interested parties simultaneously. Notably, HB
has been described as an `existential threat' by Google (the largest
beneficiary of RTB) and is at the center of an ongoing anti-trust suit brought
against Google \cite{GoogleLawsuit-2020}.  Technically, the process may occur
either through `client-side' or `server-side' HB.

\parait{Client-side HB.} When a user visits the website of a publisher, the
browser executes a script referred to as a HB wrapper. The HB wrapper solicits
bids for each ad slot from {\em all} the publisher's HB partners. These may be
DSPs, SSPs, or ad exchanges. This is done by having the browser initiate
connections with the HB servers of each partner. The requests sent to the HB
partners over these connections include information about the page and ad slot.
Since the connections are initiated by the user's browser, they also include
any cookies already set by the HB partner. Each partner may then respond to
these requests with their bids for the impression (including their ad creative
and bid value). Similar to the process for RTB, these bids are informed by any
user data available to the HB partners --- either their own or from third-party
sources.
The wrapper then forwards all bids that are received within
a pre-determined timeout period to the publisher's ad server. Finally, the
ad-server charges the winning bidder and forwards the winning ad creative to
the user's browser. Unique to client-side HB is the browser's access to all
bids, including their values and associated creatives, received for each ad
slot.

\parait{Server-side HB.} Client-side HB requires the user's browser to solicit
and forward bids from each of the publisher's HB partners. This poses a strain
on the website's performance, user's browser, and network resources.
Server-side HB provides a more efficient alternative in which the bids are
solicited directly by the publisher's (or publisher-subscribed third-party)
servers --- \ie the HB wrapper logic is moved away from the user's browser.
While this provides notable performance improvements, it poses a new challenge:
since bids are solicited from outside the user's browser, they no longer
contain the cookies set by the HB partner. This reduces the data
available to HB partners when determining the creatives and values for their
bids. To compensate for this loss of data, several server-side HB vendors
are now providing cookie syncing mechanisms so bidders may still access their
cookies prior to bidding \cite{SSHB-Prebid}.

It is also to be noted that an entity can play the role of an SSP, DSP, 
Ad exchange, third-party tracker or any combination of the four \eg Criteo is 
a notable tracker and also provides DSP services.

\subsection{Data harvesting and dissemination} \label{sec:background:data}

As is clear from the online advertising mechanisms described in
\Cref{sec:background:bidding}, user data plays a key role in determining
advertisers' bids for ad slots. With high quality data about the user, their
bids will more accurately reflect the `true value' of the impression for their
clients. This has spurred an entire industry that is focused on harvesting and
disseminating user data so that they may be sold to advertisers seeking to
optimize their bids. In this section, we provide an overview of the key methods
by which user data is harvested and disseminated.

\para{Data harvesting: stateful and stateless tracking.} Technologies to track
the online activities of users are widespread across online services and
platforms. Organizations that provide these technologies facilitate the
monetization of online services and, in return, obtain the ability to track
users across a variety of services and platforms.
Tracking technologies may be stateful or stateless. Stateful tracking
technologies rely on assigning unique identifiers to each user they encounter.
These unique identifiers are stored in the user's browser in the form of
a cookie. Therefore, when a user visits a service in which the tracking
organization is integrated, this cookie notifies the organization of the visit.
While stateful tracking presents a deterministic way of identifying users
across the web, they can be cleared by the user (making tracking impossible).
In order to address the limitations of cookies, stateless tracking mechanisms
are used. Stateless tracking does not rely on state saved on the user's device.
Instead, they work on the premise that each device is unique and relying on
real-time measurements of user's device is sufficient to identify individual
user's across services. On the Web, these measurements seek to find unique
characteristics of the user's browser using the JavaScript API provided by the
browser. These characteristics include the browser's user agent, default
languages and encoding, installed plugins, and canvas fingerprints amongst many
others \cite{Laperdrix-TWEB}. Unlike stateful techniques, these are not
deterministic --- \ie there is no guarantee that two users cannot share the
exact same characteristics. However, they are difficult to differentiate from
non-tracking uses of the browser's API --- making them harder to block.
Typically, online trackers use a combination of stateless and stateful
approaches to track users.

\para{Client-side data sharing: cookie syncing.} In order to learn of a user's
visit to a website, a tracker needs to be integrated into the website by the
publisher. Cookie syncing allows collaboration between trackers so that they
can learn of the user's visit as long as any one of them is integrated ---
essentially allowing them to share data about the user's online activities. 
This is achieved by having one partner redirect the user's browser to the
other partner by requesting for resources from it (\eg pixel images). For
example, if {\tt tracker-1.com} and {\tt tracker-2.com} are partners, {\tt
tracker-1.com} will invoke a request to {\tt tracker-2.com} causing the browser
to also send {\tt tracker-2.com}'s cookie in this request.
Cookie syncing can also be used to help data sharing partners coordinate their
databases. For example, {\tt tracker-1.com} may also provide its own unique
identifier for the user (as a parameter in the URL) in the resource request
sent by the browser to {\tt tracker-2.com}. When this request is received by
{\tt tracker-2.com}, it builds an association with {\tt tracker-1.com}'s unique
identifier and its own cookie. This allows {\tt tracker-2.com} to integrate any
out-of-band data received from {\tt tracker-1.com} into its own database
\cite{Papadopoulos-WWW2019}.
Cookie syncing is heavily used in RTB and HB to allow DSPs, who may not be
integrated on the publisher site, to make informed assessments of the
impression opportunities. Amongst the popular data sharing mechanisms in the
online advertising ecosystem, only cookie syncing is visible to the user's
browser.

\para{Out-of-band data sharing.} Cookie syncing presents two main drawbacks:
(1) they reduce the performance of webpage loads by forcing a large number of
browser redirects and (2) they are impacted by the current restrictions on
third-party cookies implemented by popular web browsers
\cite{CookiesNoLonger-Chromium2020}. To circumvent these challenges, data
sharing between entities in the advertising ecosystem is increasingly being
facilitated through mechanisms that do not involve the user's browser ---
making them invisible to users, auditors, and researchers. 

\parait{Universal ID.} Universal ID based approaches are a solution to the
recent browser restrictions on setting and accessing third-party cookies
\cite{UnifiedID, UnifiedID-TradeDesk}. The idea behind them is to leverage user
identifiers such as email addresses or user ids that are supplied to publishers
as a way to uniquely identify users across the entire advertising pipeline.
This erases the need for matching cookies in order to sync databases. At
a high-level, the process works as follows: (1) a user visits {\tt website.com}
in which {\tt tracker-1.com} is included, (2) {\tt website.com} passes the user
identifier known to it (\eg email address used during sign up) in the request
sent to {\tt tracker-1.com}, (3) {\tt tracker-1.com} records the user
identifiers presence on {\tt website.com}. When {\tt tracker-1.com} and {\tt
tracker-2.com} wish to collaborate, they can simply perform a join operation on
their databases to augment their records of each user. The most popular
implementation of universal ID is UnifiedID 2.0 by TradeDesk (endorsed by the
IAB) \cite{UnifiedID}.

\parait{Data brokering.} Data brokers are aggregators of user data from many
sources including online trackers, credit card companies, state and federal
authorities, etc. Comprehensive information about users meeting specific 
characteristics (\eg living in a specific zip code and of a specific age with a recent purchase of
a specific product) are often purchased by advertisers and synced with their
own datasets to facilitate more accurate bids in the behavioral advertising
ecosystem \cite{Venkatadri-WWW2019}.

In this paper, we propose and validate a technique to identify
tracker-advertiser data sharing relationships that: (1) is agnostic to 
ad bidding and delivery mechanism (\Cref{sec:background:bidding}) and (2)
is independent of the methods used by trackers and advertisers to share their data (\Cref{sec:background:data}). 

\section{Ad creatives and user interests}
\label{sec:interests}

In this section, we focus on testing the following hypothesis:
\textit{H1. Characteristics of ad creatives are dependent on
user interests.} 
If valid, this will: (1) demonstrate that displayed ads are dependent
on user behavior, (2) show that our methods for analyzing ad creatives are able
to capture these dependencies, and (3) satisfy a necessary condition for our
attempt to use characteristics of displayed ads to infer data sharing
relationships.
We provide a description of our methodology in \Cref{sec:interests:methods} and
our results in \Cref{sec:interests:results}.

\subsection{Methodology}\label{sec:interests:methods}

\para{Overview.} Our goal is to understand if the characteristics of ad
creatives are dependent on user interests. At a high-level, we achieve this by:
(\Cref{sec:interests:methods:personas}) creating `interest groups' (\eg arts
and sports) and curating a list of websites associated with each of these
interests;
(\Cref{sec:interests:methods:crawling}) constructing a set online personas
that crawl websites in order to signal specific interests to the advertising
ecosystem;
(\Cref{sec:interests:methods:gathering}) gathering and extracting
characteristics from the ad creatives displayed to each of these personas when
they visit a set of websites; and 
(\Cref{sec:interests:methods:analysis}) conducting statistical testing to
analyze the dependence of the extracted characteristics of ad creatives on the
interest categories of their personas.

\subsubsection{Creating interest groups}
\label{sec:interests:methods:personas}
We aim to communicate an `interest' to the advertising ecosystem solely 
by simulating browsing activity. This decision is motivated by similar previous 
work by Cook \etal \cite{Cook-PETS2020}. We start by creating 16 `interest groups' based on the 
persona categories from \cite{Cook-PETS2020} (Adult, Art, Business, Computers, 
Games, Health, Home, Kids, News, Recreation, Reference, Regional, Science, 
Shopping, Society, Sports).
To assign websites to each group, we obtain a set of popular (US Top 100)
websites belonging to each of our interest groups from Similar Web
\cite{SimilarWeb} and Alexa Top Sites \cite{AlexaRanking}. Next, to establish 
the uniqueness of each `interest group', we filter out groups that have 
significant (50\%) inter overlap. We further filter out websites 
with `cookie banners' (in accordance with GDPR \cite{GDPR}), as instrumenting 
opt-in/out was out of the scope of this work. We then manually curate the 
remaining six `interest groups' (adult, games, health, news, sports, and travel) 
and their lists, to ensure the website's fit with the corresponding interest 
group. Following the best practices highlighted in \cite{Scheitle-IMC2018}, 
our set of websites in each category were gathered in Oct-2021.

Next, we used OpenWPM \cite{Englehardt-CCS2016} to crawl each remaining site to
verify that they were functional and contained trackers on them. Verifying
tracker presence is crucial since having no trackers on a website prevents our
persona's `interest' signal from being communicated to advertisers. Tracker
presence was measured by counting the number of unique matches between web
requests generated from the page load and the EasyPrivacy tracker list 
\cite{ELEP}.
Finally, we selected 50 sites for each interest group. 
These were the 50 sites with the largest number of trackers present 
on them. The top 5 sites amongst these were used for ad collection in
\Cref{sec:interests:methods:gathering} while the remaining 45 were used to construct 
personas. One site from each group was also manually selected to communicate 
`intent' --- \ie perform an action to signify high interest in the topic 
(\eg adding an interest-related product to a shopping cart or performing a 
search about an interest-related topic). 

\subsubsection{Constructing online personas to signal interests}
\label{sec:interests:methods:crawling}
In order to signal interests to entities in the advertising ecosystem, we
constructed a total of 5400 personas --- 900 for each of our six interest
groups. Of these, half were selected to communicate intent on the `intent site'
for their associated interest group.
Each persona was associated with a unique browser running on an isolated
virtual machine with a unique IP address. This was done to ensure that tracking
entities would not misconstrue the uniqueness of each persona. The browser of
each persona was automated using OpenWPM and crawled the curated set of
websites associated with their interest groups. Following the best practices
for crawling studies \cite{Ahmad-WWW2020}, \cite{Jueckstock-WWW2021}, \cite{demir22}, our OpenWPM configuration enabled
bot mitigation and disabled tracking protection. 

\subsubsection{Gathering ads and extracting characteristics}
\label{sec:interests:methods:gathering}
\parait{Gathering ads.}
After a persona finished crawling the list of sites in its interest group,
it waited for an hour before visiting a set of ad collection sites from which
all ads were gathered.
The pause was to ensure that any signals measured by trackers were eventually
delivered to advertisers. 
The set of ad collection sites included five websites for each interest group as mentioned in 
\Cref{sec:interests:methods:personas}. 
In addition, we also gathered ads shown to a {\em control persona} with no
previous browsing history (\ie using a fresh browser and IP address). We assigned one 
website from each interest groups ad collection websites to the control group. 
We did this to remove any bias towards a single interest group. The ads
gathered by this persona serve as our {\em control group} for comparisons.
In order to gather ads, we extracted all response URLs containing images and
matched them with the set of filters from EasyList. The
matching URLs denote known advertising domains that sent an image. The images
associated with each of these domains were then filtered to remove any images
smaller than 20KB (to remove icons or pixel images). The remaining images were
the ad creatives delivered to each of our personas. Manual validation was
performed on a random subset of 300 unique images to verify that the remaining
images were ad creatives. In total, our personas gathered {5.3M} ads of
which {31.5K} were unique. 
Importantly, {\em our approach gathers all ad creatives regardless of whether
they were associated with RTB or HB}.

\parait{Crawl synchronization.}
We executed our crawls in nine serial {\em runs} spanning over a 45 day period.
Each run consisted of the execution of 100 instances of each persona.
Our crawls were executed to ensure that the start of each of our nine runs were
synchronized across the six interest groups --- \ie the $n^{th}$ run of all
interest groups began simultaneously. Each run was executed only after
completion of the prior run. This precaution was taken as a best-effort attempt
to mitigate the effect of any latent temporal confounders that might impact our
subsequent data collection.

\parait{Extracting ad characteristics.}
To extract characteristics of each ad, we relied on Google Cloud Platform's (GCP)
Vision API \cite{GCPVisionAPI}. Specifically, for each recorded ad creative, we
used it to obtain the following characteristics: (1) all written text from the
image and (2) textual descriptions of identified landmarks and logos in the
image. Taken all together, these extracted characteristics present a textual
description of the supplied ad creative. 

\subsubsection{Analyzing ad dependence on interest groups}
\label{sec:interests:methods:analysis}
\parait{Converting ad descriptions to count vectors.}
The extracted characteristics of each ad creative effectively act as its
semantic textual description, allowing us to use standard text analysis
techniques to measure semantic similarity between ad creatives. 
We aggregated the descriptions of ads into nine `documents' 
for each interest group (and control group). Each document consisted of the descriptions of ads
shown to the 100 personas belonging to one run and one interest group. This
aggregation was necessary to deal with sparsity of keywords and facilitate
significance testing. 
We then created a global corpus consisting of all words occurring in all
documents. Finally, each document $d$ was converted into a count vector $X^d$
where $x^d_i$ denoted the frequency of the $i^{th}$ word (from the global
corpus) in document $d$. At the end of this step, we were left with nine 
count vectors for each interest and control group.

\parait{Measuring dependence on interest group.}
We computed the cosine similarity between the count vectors of each pair of
documents. This presented us with: (1) a distribution of within-group
similarities --- \ie a distribution representing the measure of similarity
of ads shown to the same interest group across different periods of time and
(2) a distribution of across-group similarities for each pair of interest
groups --- \ie a distribution representing the measure of similarity of ads
shown to the two interest groups. 
Finally, to validate our hypothesis about the dependence of ads on interest
groups, we analyzed the mean values of each of these similarity distributions and
tested their differences for statistical significance using a two-sample
$t$-test ($p$ < .05). 
Put another way, for any pair of interest groups ($g_1$, $g_2$), we compare
$D_{sim}(g_1, g_2)$ and $D_{sim}(g_1, g_1)$ using a two-sample $t$-test. Here,
$D_{sim}(g_i, g_j)$ denotes the distribution of cosine similarities between
the count vectors of $g_i$ and $g_j$.
We conclude that our hypothesis is valid if our findings indicate: (1) high
within-group similarities --- \ie semantics of ads shown to an interest group
are similar over time, (2) low across-group similarities --- \ie semantics of
ads shown to two unrelated interest groups are not similar, and (3) statistical
significance between the differences in within- and across-group similarities.

\subsection{Methodology validation and results} 
\label{sec:interests:results}

We now present the results of our experiments to validate our methodological
decisions and our hypothesis. Specifically, we:
(\Cref{sec:interests:results:nsites}) test the impact of number of
interest-related sites crawled by a persona on the number of ads received;
(\Cref{sec:interests:results:ips}) validate our decision to use unique VMs and
IP addresses for each persona;
(\Cref{sec:interests:results:extracting}) validate the quality of ad
descriptions returned by the GCP Vision API;
(\Cref{sec:interests:results:dependence}) demonstrate the dependence of ad
creatives shown to our personas on the personas' interest group; and
(\Cref{sec:interests:results:intent}) examine the impact of demonstrating
`intent' during crawling.

\subsubsection{Validation: Number of sites in an interest group}
\label{sec:interests:results:nsites}
Advertisers are likely to make stronger bids (higher than the floor value of
the slot)  on users for whom they  have strong signals of a specific interest.
Consequently, one expects this to result in more targeted ads and less unfilled
ad slots --- crucial to the success of our study. 
However, it is unclear {\em how many websites to include in each interest group
for our personas to signal a strong and specific interest}. Although we expect
that more websites related to one interest will always be better, the scale of
our measurements require us to find a workable trade-off between practicality
and signal strength.
To address this question, we conducted a pilot experiment in which the number
of websites in each interest group was varied and the number of ads and unique
ads gathered from our ad gathering sites were measured. 
In this experiment, we altered the number of websites in each interest group
from 5 to 50 (in increments of five) using the same website selection process
outlined in \Cref{sec:interests:methods:personas}. Then, we created 40 personas
for each of these interest groups using the same approach outlined in
\Cref{sec:interests:methods:crawling}. Finally, we measured the total and
unique number of ads displayed to each of these personas on our set of ad
gathering sites. 
%
%
We found a steady increase in the median number of unique ads presented to each
persona until the size of our interest groups reached 40 websites. Increasing
the number of websites in the interest groups beyond 40 websites resulted
in marginal and strongly diminishing gains. These results informed our final
decision to use 45 websites in each of our interest groups. 

\subsubsection{Validation: Unique IPs and VMs}
\label{sec:interests:results:ips}

Allocating personas to their own VM and IP address for a large-scale study such
as ours is monetarily expensive and challenging to automate. To decide whether
this effort was necessary, we conducted a pilot experiment to determine whether
such `sandboxing' of personas impacted the uniqueness of ads received by them.
In this experiment, we selected three interest groups (adult, sports, and
travel) containing 45 websites and allocated 100 personas to each. We allocated
unique VMs and IP addresses to 50 of the personas of each interest group. The
remaining set of 150 personas (50 from each interest group) used the same VM
and IP address. 
For each setting (isolated and non-isolated), we: (1)  extracted all the
unique ads shown to each interest group (\ie we do not count the same ad shown
twice to a persona) and (2) measured the fraction of these ads that never
occurred on any of the other interest groups (\eg the fraction of unique ads
shown only to our sports personas). Observing a difference between these
fractions measured from each setting would suggest that isolating a persona
influences the uniqueness of the ads received by it. 
Our results are shown in \Cref{tab:interests:results:ips}. We see a notable
difference in the fraction of ads unique to each interest group ---
particularly in the travel interest group. To verify the significance of these
differences, we used a $\chi^2$-test to test the dependence of number of ads
unique to each interest group on the isolation setting used. We found
a statistically significant relationship between these variables ($p$ < .05).
There have been numerous research efforts to highlight the importance of crawl
configuration decisions on inferred results \cite{Jueckstock-WWW2021,
Ahmad-WWW2020}. Our work uses the best practices highlighted in these studies
and also contributes to them. One insight from this pilot experiment is the
need to create `sandboxed' personas during measurements of personalized ads.  
This result informed our decision to isolate each of the personas used in our
study.

\begin{table}
	\centering
	\begin{tabular}{c cc}
    {\bf Interest Group}  & \multicolumn{2}{c}{\bf Ads unique to interest group} \\
                          & Isolated  & Non-isolated  \\
    \midrule
    Adult                 &  51\% & 43\% \\
    Sports                &  42\% & 48\% \\
    Travel                &  72\% & 49\% \\
	\end{tabular}
    \caption{Impact of using isolated VMs and IP addresses on the fraction of
    ads that are unique to each persona ({\em cf.}
    \Cref{sec:interests:results:ips}).
    }
    \label{tab:interests:results:ips}
\end{table}

\subsubsection{Validation: Quality of ad descriptions}
\label{sec:interests:results:extracting}
The Google Vision API provides us with the capabilities to extract features
such as logos, text, faces, landmarks, objects, and web entities present in an
image. We conducted a pilot experiment to determine which of these features
would facilitate meaningful text descriptions that captured the semantics of
the ad. We randomly sampled 263 unique ads obtained from our prior experiment
(\Cref{sec:interests:results:ips}) and manually analyzed their images and the
corresponding data extracted by the Vision API.
The API identified text in 86\%, logos in 28\%, and faces in 16\% of all
images. Upon manual inspection, we found the text extracted from these images
to be 100\% accurate. Perfect accuracy was also found for logos extracted from
these images when the API returned a confidence score greater than 80\%. Other
extracted features were found to be limited in their confidence scores and
accuracy. 
Based on these results, we exclusively relied on the text and high-confidence
brand/logo description features extracted from each image.

\subsubsection{Dependence of ad creatives on interests}
\label{sec:interests:results:dependence}

Using the methodology outlined in \Cref{sec:interests:methods}, we analyzed
the dependence of our features extracted from ad creatives on the interest
groups they were served to. 
If our hypothesis is valid and our method for extracting features from ad
creatives is accurate, we expect to find that: (1) there are high similarities
in the ads shown to personas belonging to the same interest groups; (2) there
are low similarities in the ads shown to personas belonging to different
interest groups; and (3) these differences in similarities are statistically
significant.

\begin{figure}[t]
  \includegraphics[width=0.5\textwidth]{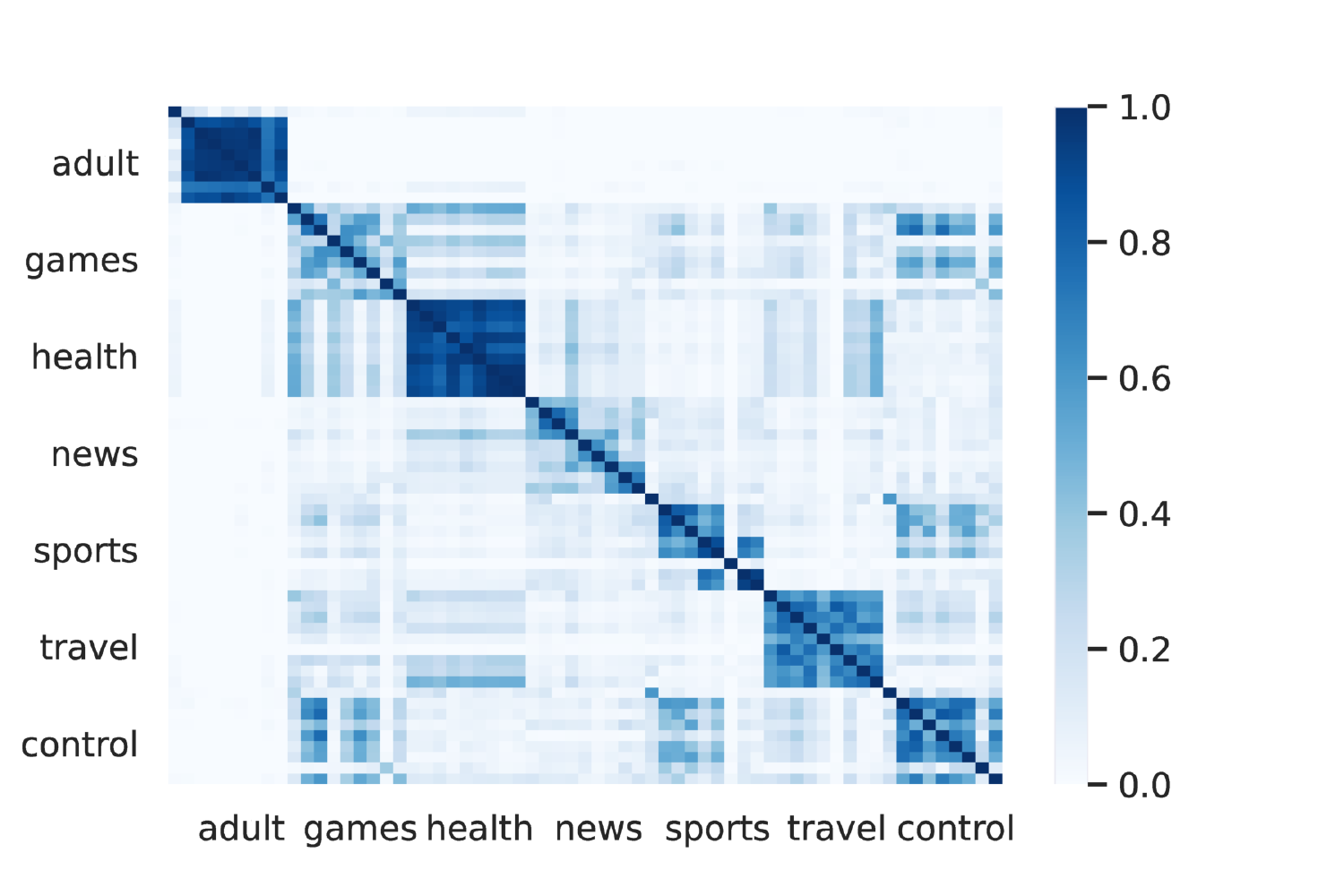}
  \caption{\label{fig:interests:results:dependence} Heatmap depicting the
  inter- and intra-group cosine similarities across the nine runs and interest
  and groups.}
\end{figure}

\parait{Within- and across-run similarities.}
\Cref{fig:interests:results:dependence} visualizes the similarities of features
extracted from ad creatives within and across each run and each interest group.
Each interest group contains nine runs and each cell in the heat map represents
the cosine similarity between the count vectors obtained from the corresponding
runs of the associated interest groups. 
We notice a pattern of higher similarity scores along the diagonal. This
indicates that ads shown to the same interest group are in fact more similar
that ads shown to different interest groups. 
Importantly, for every run, we find that the mean within-group similarities
are significantly higher than the mean across-group similarities --- providing
evidence that ads are influenced by our interest groups and personas' browsing
histories. This difference is particularly high in the adult and health
interest groups which suggests that these interest groups are the subject of
significantly different ad targeting.

\parait{Significance testing of differences in within- and across- interest
group similarity distributions.} 
Following the procedure outlined in \Cref{sec:interests:methods:analysis}, for
each pair of interest groups ($g_1, g_2$), we verified the statistical
significance of differences in the within- and across-group similarity
distributions (\ie $D_{sim}(g_1, g_1)$ and $D_{sim}(g_1, g_2)$, respectively).
We found that all pairs of interest groups where $g_1 \neq g_2$ had
statistically significant differences (two-sample $t$-test with $p$ < .05).
\Cref{tab:interests:results:dependence} shows the means of each of these
distributions --- \ie $D_{sim}(g_i, g_j)$. These results {\em validate our
hypothesis that the characteristics of ad creatives are dependent on user
interests and our methods to extract these characteristics.}

\begin{table*}[t]
\begin{center}
\begin{tabular}{cccccccc} 
  & \textbf{Adult} & \textbf{Games} & \textbf{Health} & \textbf{News} & \textbf{Sports} & \textbf{Travel} & \textbf{Control} \\ 
\midrule

  \textbf{Adult}     & 0.71  & 0.00 & 0.01 & 0.00  & 0.00  & 0.00 & 0.00 \\

  \textbf{Games}     & 0.00  & 0.42 & 0.19 & 0.07 $^\uparrow$  & 0.13 & 0.13 & 0.29 \\

  \textbf{Health}    & 0.01  & 0.19 & 0.87 $^\downarrow$ & 0.13 & 0.04 & 0.18 & 0.07 \\

  \textbf{News}      & 0.00 & 0.07 $^\downarrow$ & 0.13 & 0.36  & 0.09 & 0.05 & 0.07 \\

  \textbf{Sports}    & 0.00  & 0.13 & 0.03  & 0.09 & 0.34 & 0.05 $^\uparrow$  & 0.22 \\

  \textbf{Travel}    & 0.00  & 0.13 & 0.18  & 0.05 & 0.05 $^\uparrow$ & 0.62  & 0.10 \\

  \textbf{Control}   & 0.00  & 0.29 & 0.08  & 0.08 & 0.22 & 0.10 & 0.41 \\
\end{tabular}
\end{center}
  \caption{Mean of the distribution of ad similarities between each pair of
  interest groups. 
  Each cell associated with the interest groups $g_1$ and $g_2$ denotes the
  mean of the distribution of similarities of ads shown to personas from $g_1$
  and $g_2$.
  Higher values indicate higher similarity between the ads for the
  corresponding interest profiles. 
  In all cases where $g_1 \neq g_2$, the differences in the distributions of
  within- and across-group similarities was statistically significant
  ({\em cf.} \Cref{sec:interests:results:dependence}).
  %
  %
  $\uparrow$ and $\downarrow$ indicate a statistically significant increase or
  decrease in this similarity when no `intent' action was communicated ({\em
  cf.} \Cref{sec:interests:results:intent}).
  \label{tab:interests:results:dependence}
  }
\end{table*}

\subsubsection{Influence of communicating intent}
\label{sec:interests:results:intent}

Previous work has shown that communicating intent impacts the behavior of
online advertisers. Specifically, Cook \etal \cite{Cook-PETS2020} showed that
communicating intent caused advertisers to increase the value of their bids. We
verified whether this also caused a change in the ads displayed to
users. 
To measure the impact of intent, we compared the within- and across-group
similarity distributions for the `intent' and `no-intent' personas --- \ie for
all pairs ($g_1$, $g_2$), we compare $D_{sim}(g_1^{\text{intent}},
g_2^{\text{intent}})$ with $D_{sim}(g_1^{\text{no intent}}, g_2^{\text{no
intent}})$ using a two-sample $t$-test ($p$ < .05).
Our results are illustrated in \Cref{tab:interests:results:dependence}. Here,
$\downarrow$ and $\uparrow$ denote as statistically significant decrease and
increase, respectively, in the mean of the corresponding $D_{sim}(g_1, g_2)$
distribution when switching from intent personas to no intent personas.

\parait{Within-group similarities.}
When $g_1 = g_2$, a statistically significant decrease ($\downarrow$) suggests
that ads displayed to a personas in $g_1$ become more diverse when intent
is communicated during crawling. This case occurred only for personas in our
health interest group. No significant increases were found when $g_1 = g_2$.

\parait{Across-group similarities.}
When $g_1 \neq g_2$, a statistically significant decrease ($\downarrow$)
suggests that the ads displayed to $g_1$ and $g_2$ became more similar when
intent was communicated. This case occurred for the (news, games) interest
group pair. 
A statistically significant increase ($\uparrow$) indicates that the
ads displayed to $g_1$ and $g_2$ became less similar when intent was
communicated. This case occurred for the (sports, travel) interest group pair.
In general, our results {\em do not suggest a widespread influence of intent on
ad creatives delivered to personas.}

\subsubsection{Takeaways}
\label{sec:interests:results:takeaways}
Taken all together, our analysis allows us to
conclude that {\em characteristics of ad creatives are dependent on user
interests} and that {\em our methods for communicating user interests and
characteristics of ad creatives are effective}.
Further, we also find that:
a browsing history of about 40 sites related to a specific interest is
sufficient for measuring personalized ads related to that interest
(\Cref{sec:interests:results:nsites}); 
isolating unique personas so that they possess their own unique browser, VM,
and IP address has a significant influence on their measured personalized ads
(\Cref{sec:interests:results:ips}); 
extracting text and logo descriptions are sufficient to obtain an effective
text representation of ad creatives (\Cref{sec:interests:results:extracting});
and 
communicating `intent' during a crawl does not have a widespread influence on
the nature of personalized ads received when interests are already communicated
by long and interest-specific browsing histories
(\Cref{sec:interests:results:intent}).
Additional configuration details, websites used for personas, and ads gathered can
be found at \atomarchive.

\section{Inferring data sharing }
\label{sec:relationships}

\begin{figure*}[t]
  \centering\includegraphics[width=0.9\textwidth]{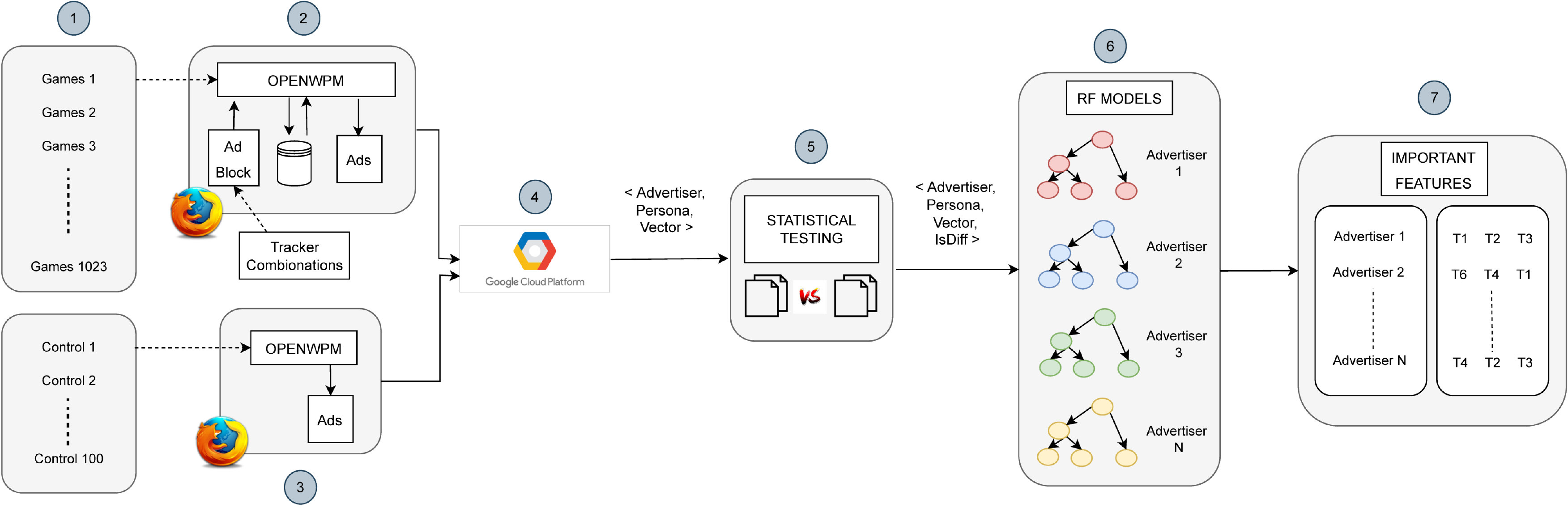}
  \caption{\label{fig:relationships:methods:figure} Overview of ATOM. 
  \textcircled{1} We curate websites associated with interest-related and control personas. 
  \textcircled{2} Using OpenWPM, we train and log ads shown to each
  interest-related persona while blocking some combination of tracking
  organizations. 
  \textcircled{3} Simultaneously, we collect ads observed by each control persona.
  \textcircled{4} We extract labels from gathered ads.
  \textcircled{5} We measure significance of differences in labels extracted
  from ads shown by individual advertiser to our control and interest-related
  personas.
  \textcircled{6} We use Random Forest models to fit the targeting behavior of
  each advertiser.
  \textcircled{7} We extract trackers found to have a high influence on the
  accuracy of the Random forest model associated with each advertiser.   }
\end{figure*}

Our previous results effectively show that it is possible to use features
extracted from ads to measure the relationship between interest groups and
personalized ads.
We now use the insight that {\em specific trackers are responsible for
communicating these interest groups to the advertisers of these personalized
ads}. Therefore, by systematically blocking specific trackers, we can use these
ad features to understand which advertisers are no longer able to deliver ads
similar to those typically associated with the interest group --- signifying
the presence of a data sharing relationship between the set of blocked trackers
and the advertiser.
We operationalize this insight to build ATOM. ATOM is a generalizable framework
to identify evidence of tracker-advertiser data sharing relationships. By
providing validation of ATOM's inferred data sharing relationships, we also
validate the hypothesis {\em H2. Characteristics of ad creatives can be used to
infer tracker-advertiser data sharing relationships}.
We provide a description of the methods used to construct and empirically
evaluate ATOM in \Cref{sec:relationships:methods} and highlight the results
from a test deployment in \Cref{sec:relationships:results}. \Cref{fig:relationships:methods:figure} 
gives a high-level overview of ATOM's architecture. 

\subsection{Methodology}

\label{sec:relationships:methods}

\para{Overview.} Our goal is to use the features of personalized ads to
infer-tracker advertiser relationships. We achieve this in two distinct phases:
a data collection phase in which online personas are used to gather ads while
selectively blocking trackers of interest and a relationship inference phase
in which the influence of specific trackers are measured on the ads delivered
by each advertiser.
In our data collection phase,
(\Cref{sec:relationships:methods:crawls}) we configure ATOM to conduct crawls
while selectively communicating interests to a specific set of trackers; and
(\Cref{sec:relationships:methods:ads}) then we gather, label, and identify the
advertiser associated with the ads observed after ATOM's crawls are complete. 
Once our data collection is complete,
(\Cref{sec:relationships:methods:changes}) we use a statistical approach to
determine which blocking conditions caused which advertisers to change
their behavior; and
(\Cref{sec:relationships:methods:learning}) account for the probabilistic
nature of data sharing and multi-tracker data sharing relationships to finally
produce inferences about the trackers responsible for enabling the
personalization capabilities of observed advertisers.

\subsubsection{Data collection: Crawling to signal interests}
\label{sec:relationships:methods:crawls}
Our goal in this stage is to conduct crawls that signal an interest to
a specific set of trackers. We use lessons from our previous analyses
to inform our crawling methods. 

\parait{Selecting an interest group.} We create {\em one} interest group with
45 manually curated websites ({\em cf.} \Cref{sec:interests:results:nsites}).
It is important that: (1) the trackers whose data sharing relationships are
being studied are highly present in the websites associated with the selected
interest group and (2) a large number of advertisers are seeking to deliver
personalized ads to personas in this interest group. 

\parait{Constructing online personas and controlling interest leakage.}
Next, we configure ATOM to create online personas that selectively hide their
interests from a specific set of trackers. 
Given a set of trackers, ATOM creates one persona for which all communication
to that set of trackers is blocked by the browser. No other tracker from 
any organization is blocked in this step. This is done by matching the
browser's outgoing requests with the set of filter rules associated with the
supplied set of trackers. Each of these created personas then begin to crawl the 
websites in their associated interest group.
Based on our results in \Cref{sec:interests:results:ips}, ATOM associates
a unique VM instance and IP address to each created persona.

\subsubsection{Data collection: Collecting and processing ads}
\label{sec:relationships:methods:ads}
Once all the `interest signaling' crawls are complete, ATOM pauses for a period
of two hours. This is to ensure that any interest signals gathered by trackers
are disseminated through the advertising ecosystem. The duration of our wait 
period was influenced by prior work by Cook \etal \cite{Cook-PETS2020}.
Next, each persona visits a pre-determined set of websites to gather and
extract features from all displayed ads using the same process described in
\Cref{sec:interests:methods:gathering}. During this phase of data gathering,
all trackers are unblocked for all personas to prevent accidental blocking of
any ad delivery mechanisms. 

\parait{Identifying advertisers associated with displayed ads.}
Unlike our experiments in \Cref{sec:interests}, it is now necessary to identify
the advertiser (typically, the DSP) associated with each ad. 
This is done by identifying the response (sent to the browser) that is
responsible for delivering the ad creative. The domain of the source of this
response is extracted and its parent organization is identified from a supplied
list of domain-organization mappings. This parent organization is labeled as
the advertiser associated with this ad creative.

\parait{Collating data.}
Finally, ATOM combines (by summing together) all the count vectors associated
with ads delivered by each advertiser to each persona (represented as the set
of trackers blocked in its configuration). This produces a count vector record
for each advertiser-persona pair. These {\em vector records} are saved as:
{[\tt advertiser, persona, <combined count vector>]}.

\subsubsection{Analysis: Finding changes in advertiser behavior}
\label{sec:relationships:methods:changes}

Once data collection is complete, ATOM possesses multiple sets of (advertiser,
persona) vector records --- each associated with one data collection run. Next,
it uses these records to identify all the (advertiser, persona) pairs that
demonstrated a change in the advertiser's behavior.
This is done by using the (advertiser, control) vector records as a reference
for the advertiser's typical behavior when no blocking is performed. 
A $\chi^2$-test is used to identify the (in)dependence between the observed
count vectors and their sources (control or persona). If a dependence is found
($p$ < .05), the advertiser is identified to have sent statistically different
ads to the control and the persona (which is associated with a specific
blocking condition). This indicates a change in behavior caused by the blocking
of the specific trackers.
Following this step, each vector record is augmented with its significance
status with reference to the control and are of the form {\tt [advertiser,
persona, <combined count vector>, is\_different\_from\_control]}.

\subsubsection{Analysis: Inferring data sharing relationships}
\label{sec:relationships:methods:learning}

\parait{Why relationship inferences cannot be deterministically `solved'. }
In a noise free and deterministic scenario, the vector records from the prior
step would be sufficient to `solve for' all tracker-advertiser data sharing
relationships --- one would only need to find the minimal set of tracker
blocking configurations under which an advertiser demonstrated a statistical
difference from the control.
Unfortunately, this is not likely because online advertising is an inherently
probabilistic process in which advertisers may simultaneously lose one ad
auction and win another with identical bid values, users, and user data. This
can be for reasons including minor network delays that prevent a bid response
from reaching the auction on time, a change in strategy by other bidders, or
even restrictions on the rate at which ad inventory can be used.
The presence of this noise also impacts our data --- \eg it may create cases
where a tracker-advertiser relationship is present but not shown in our records
because the corresponding advertiser did not win enough auctions for us to
extract meaningful features from their creatives. 
Further complicating analysis are the facts that: (1) it is practically
impossible to guarantee that all trackers belonging to a specific organization
are blocked and (2) many advertisers may have relationships with several
tracking organizations --- therefore, it is possible that even when $n-1$ of
these organizations are accounted for and blocked the $n^{th}$ tracker
organization is able to observe and communicate persona interests with the
advertiser.

\parait{Inferring relationships with interpretable statistical models.}
To account for such noise, similar to prior work \cite{Cook-PETS2020}, ATOM
uses an interpretable random forest model to learn the correlations between
the presence/absence of a tracker and the changes in
advertiser behavior. For each advertiser, this is done as follows:

\begin{itemize}

  \item {\it Segmenting vector records.} ATOM splits the multiple vector
    records {\em for each} (advertiser, persona) pair into a cross-validation
    set and a holdout set. For example, if there are ten
    ($\text{advertiser}_i$, $\text{persona}_j$) records --- one from each of
    ten data collection runs, a sample of eight may be placed in
    a cross-validation set and the remaining two may be placed in a holdout
    set. 

  \item {\it Model building and cross-validation.} The vector records in the
    cross-validation dataset are split into a predetermined number of {\em
    folds} such that each fold contains the same number of records associated
    with each persona. 
    We then construct a random forest model that uses the tracker blocking
    configuration (obtained from the {\tt persona id}) as a feature and seeks
    to predict whether {\tt is\_different\_from\_control} is {\tt True} or {\tt
    False}.
    Models are built using a grid search over a set of random forest
    configuration parameters and the model with the highest average accuracy
    over all the cross-validation folds is returned.

  \item {\it Model testing.} The best performing cross-validated model is then
    presented with the features from the holdout set and its accuracy is
    measured. 

  \item {\it Relationship inference.} If a model demonstrates reasonably high
    accuracy (determined by a supplied threshold) over the holdout set, it has
    effectively learned of correlations
    between tracker presence and the advertiser's behavior. When this occurs,
    ATOM extracts the information gain associated with each feature (\ie
    tracker presence) from the model. This is a measure of the feature's
    importance in aiding the model's accuracy. ATOM makes an inference of
    a data sharing relationship between an advertiser and a tracker if the
    information gain associated with the tracker is more than one standard
    deviation higher than the mean information gain for all features. Note that
    this is a conservative approach to reduce the likelihood of
    false-positives.

\end{itemize}

The decision to use a random forest model was due to its interpretability and
lower susceptibility to overfitting. The decision to use a holdout set for
testing was to ensure that overfitting did not occur.

\subsubsection{Configuration for ATOM's test deployment}
\label{sec:relationships:methods:atom}

\parait{Data collection configuration.}
Based on the data obtained from our analysis in \Cref{sec:interests:results},
we selected the `games' interest group in our test deployment of ATOM. 
To select trackers to analyze, we used the following process:

\begin{itemize}
  \item {\it Obtaining a list of trackers present in the interest group.} We
    used EasyList to identify all tracking domains observed in the `games'
    interest group using our crawl data from \Cref{sec:interests}. 
  \item {\it Identifying the parent organizations of identified trackers.} We
    used external data sources including WHOIS records, TLS certificates, and
    WebXray \cite{Libert-WWW2018} to identify the parent organizations of each
    identified tracker. Organization names in WHOIS records and TLS
    certificates have been used and validated in prior work seeking to identify
    common owners of domains and network infrastructure
    \cite{Razaghpanah-NDSS2018, Bashir-IMC2019, Pouryousef-PAM2020, Mao-PA2003,
    Cho-CoNEXT2017}. This same process was also used to identify the parent
    organizations of advertisers. 
  \item {\it Selecting trackers to analyze.} Using the organization names
    obtained from this process, we grouped all our tracking domains by
    their parent organization. We selected the ten organizations with the
    largest number of trackers as the subject of our test deployment. These
    organizations were: Alphabet, Rubicon, Adobe, GumGum, OpenX, Pubmatic,
    Index Exchange, Facebook, 33Across, and Oracle.
\end{itemize}
Each persona was then associated with one unique combination of blocked
tracking organizations. This resulted in a total of 1,024 (\ie $2^{10}$)
personas. An additional 100 {\em control personas} that performed {\em no
tracker blocking} were also created. 
Finally, each persona began their data collection as described in
\Cref{sec:relationships:methods:crawls}-\Cref{sec:relationships:methods:ads}.
The process was repeated ten times over the period of two months. 

\parait{Analysis configuration.}
Our test deployment split the ten sets of vector records such that eight sets
were used for cross-validation and two sets were used for holdout testing.
A 60\% threshold was used for relationship inference --- \ie we only report
inferences from models that had higher than 60\% accuracy on the holdout data.

\subsection{Results}
\label{sec:relationships:results}

We now present the results of our test deployment of ATOM. Specifically, we:
(\Cref{sec:relationships:results:inferences}) present relationship inferences
made by ATOM using the configuration outlined in
\Cref{sec:relationships:methods:atom} and then
(\Cref{sec:relationships:results:validation}) validate these inferences using
existing techniques to identify tracker-advertising data sharing relationships.

\subsubsection{Relationships inferred by ATOM}
\label{sec:relationships:results:inferences}
\parait{Advertiser model accuracy.}
In total, the models developed by ATOM for nine advertisers were found to have
an accuracy higher than our 60\% threshold. A summary of our results for these
models is provided in \Cref{tab:relationships:results}. 
We find that the accuracy of models built to understand OpenX, EAI, and The
Trade Desk was 100\%. Further analysis shows that this occurs when a very large
fraction of tracker blocking configurations (\ie personas) result in
statistically different ads than the control --- suggesting that these
organizations have relationships with nearly all of the ten largest
tracking organizations. This is confirmed by the presence of a nearly uniform
distribution of information gain across all trackers and low standard deviation
of information gain.
In general, we find that our models outperform previous work using bid values
as a side channel.

\parait{Relationships inferred from model interpretation.} 
ATOM was able to identify 11 tracker-advertiser relationships for
nine advertisers. We note that our decision to only return the relationships
with a information gain of one standard deviation higher than the mean results
in very conservative inferences about the existence of data sharing
relationships. This was specifically chosen to reduce the rate of
false-positives from our deployment.
Of the nine advertisers, Alphabet appears as one of the most influential data
suppliers to six. This is not surprising since: (1) Alphabet trackers are the
most widespread and (2) their dominance in the RTB bidding ecosystem (as an SSP, 
adexchange, and DSP) is well known. Interestingly, from the information gain
associated with Alphabet trackers, we find that several large advertisers
including Amazon, Flashtalking, and Criteo appear to have nearly exclusive
relationships with Alphabet amongst the ten largest tracking organizations.
OpenX was the second most connected tracker with relationships to Exponential
Advertising Intelligence and Media Math.

\begin{table}[t]
  \begin{tabular}{p{.7in}cp{1.6in}} 
  {\bf Advertiser} & {\bf Accuracy } & {\bf Inferred relationships} \\ 
\midrule
    OpenX                         & 100\% & Oracle$^{1}$ (.12), Alphabet$^{2}$ (.12) \\
    Exp.Ad.Intel                  & 100\% & OpenX$^{2}$  (.27)    \\
    Trade Desk                    & 100\% & GumGum$^{2}$ (.15)  \\
    Adform                        & 99\%  & Alphabet$^{2}$ (.20) \\
    Pubmatic                      & 97\%  & Alphabet$^{2,3}$ (.15) \\
    Media Math                    & 96\%  & OpenX$^{2}$ (.17), {Facebook} (.15) \\
    Amazon                        & 88\%  & Alphabet$^{2}$ (.66) \\
    Flashtalking                  & 86\%  & {Alphabet} (.67) \\
    Criteo                        & 60\%  & {Alphabet} (.90) \\
\end{tabular}
\caption{\label{tab:relationships:results} Summary of advertiser model
  performances and inferred relationships. `Accuracy' denotes the constructed
  model's performance over the holdout set. `Inferred relationships' denotes
  the trackers identified by ATOM to have highest influence on the
  corresponding advertiser's performance and their corresponding information
  gains. $^1$, $^2$, and $^3$ denote that the
  relationship was verified using CCPA disclosure documents, analysis of
  client-side cookie syncing, and by analyzing header bidding bids,
  respectively. }
\end{table}

\subsubsection{Validation of inferred relationships}
\label{sec:relationships:results:validation}

To validate our results we use prior work (analyzing bid values and cookie
syncing) and external sources of data (CCPA public data sharing disclosures) to
identify data sharing relationships. We were able to validate nine of our 11
inferred data sharing relationships.

\parait{Validation via CCPA disclosures.} 
In accordance with the CCPA, all data brokers must disclose their data sharing
partnerships. These can be identified through analysis of their privacy
policies. 
Unfortunately, the CCPA's limited definition of a data broker does not include
the trading of de-identified user data obtained by typical online trackers.
Therefore, of all the tracking organizations in our test deployment, only
Oracle is currently registered as a data broker. We were able to verify the
relationship between Oracle and OpenX through their disclosure.

\parait{Validation via KASHF \cite{Cook-PETS2020}}. 
KASHF identifies data sharing relationships by analyzing changes in an
advertiser's bid values for a persona. This bidding behavior is visible for the
specific case where publishers facilitate client-side header bidding using {\tt
prebid.js}. 
During our experiment with ATOM, all header bidding bids were recorded during
the data collection.
We used KASHF on this dataset of bids. We were able to validate the
relationship between Alphabet and Pubmatic. No other relationships could be
identified or validated.

\parait{Validation via client-side cookie syncing.}
Finally, we validate each of our inferred data sharing partners by analyzing
our data for cookie syncing relationships between them. Cookie syncing is
identified by finding any redirect chains that contain a cookie. We use the
framework developed by Iqbal \etal \cite{Iqbal-NDSS2022} to identify cookie
syncing from our logs of HTTP requests and responses.
In total, we identified 7 advertisers engaging in cookie syncing
relationships. Of these, 4 were performing cookie syncing with an
Alphabet-owned tracker.

\subsubsection{Takeaways}
Taken all together, we conclude that the {\em characteristics of ad
creatives can be used to infer tracker-advertiser data sharing relationships}.
We prove this hypothesis by building and deploying ATOM. By only leveraging
features from personalized ads, ATOM is able to build high quality models of
several advertisers and infer their data sharing relationships with trackers
(\Cref{sec:relationships:results:inferences})
while maintaining a low false-positive rate
(\Cref{sec:relationships:results:validation}).

\section{Related work}\label{sec:related}

\para{Measurements of data gathering practices.} Significant work has been
done to catalog the data gathering practices of online advertisers and
trackers. Krishnamurthy \etal performed longitudinal measurements, using
automated browser extensions, to quantify prevalence of trackers
\cite{krishnamurthy-SIGCOMM2006, krishnamurthy-WWW2009}. They showed a 30\%
increase in tracker presence on popular websites. 
Following work by Roesner \etal \cite{Roesner-NDSI12} measured more complicated
aspects of tracking such as cookie syncing. Analyzing a wider range of tracking
mechanisms allowed them to show that 20\% of a users browsing history is
gathered by trackers. 
Whereas, works by Cahn \etal \cite{Cahn-WWW16} and Papadopoulos \etal
\cite{Papadopoulos-WWW2019} on characteristics of web cookies and cookie
syncing, established high prevalence of cross site tracking using cookies. 
Most recently, Iqbal \etal \cite{Iqbal-SP2020, Iqbal-SP2021} used machine
learning approaches to identify a variety of stateful and stateless tracking
approaches used by online trackers.
Other work has focused on the interplay between the data gathering and online
targeting ecosystems. Specifically, Olejnik \etal \cite{Olejnik-HAL13,
Olejnik-16} measured how mechanisms of the online advertising ecosystem (such
as Real Time Bidding) could be exploited to facilitate user data gathering.
Bashir \etal \cite{bashir-pets-2018} further highlighted this by empirically
demonstrating that mechanisms such as Real Time Bidding were exploited by many
online tracking entities to provide them with access to up to 92\% of a user's
browsing history.
Research contributions have also included the development of platforms and
methodologies for measuring online data gathering practices. For example,
researchers have built tools such as XRay \cite{Lecuyer-USENIX2014},
FPDetective \cite{Acar-SIGSAC13}, OpenWPM \cite{Englehardt-CCS2016}, and
AdGraph \cite{Iqbal-SP2020} to enable reliable and scalable measurements of
online tracking behaviors.

\para{Measurements of data sharing practices.} Data gathering has been
studied extensively due to its visibility in the browser.
However, it becomes extremely difficult to observe data after it exits the
browser. Furthermore, as highlighted in \Cref{sec:background:data}, advertisers
and trackers have a natural incentive to share data to maximize their
performance and revenue. This makes it crucial to understand these data flows.
To our knowledge only two works in the past attempt to address this issue.
%
Bashir \etal \cite{Bashir-Security2016} trained
personas and gathered retargeted ads to uncover data flows between trackers and
online retargeting advertisers. 
The key idea being that: If an advertiser $A$, serves a retargeted ad to
a persona without observing it directly, this behavior can be indicative of
server-side data sharing between $A$ and the trackers that observed the
persona. 
This technique sets a lower bound on identifying server side relationships as
it only considers a specific case of personalized advertising --- \ie
retargeting. Furthermore, manually identifying retargeted ads faces serious
scalability challenges. {\em In contrast with this work, ATOM programmatically
includes all categories of possible personalized ads --- not just retargeted
ads.}
%
Cook \etal \cite{Cook-PETS2020} introduced KASHF,
a programmatic framework to train personas and identify server side
relationships. Leveraging exposure to bid values from client side
Header Bidding, KASHF measures correlation between presence of a tracker and bid
values. If blocking a tracker during persona training alters bidding patterns
of an advertiser, they conclude there exists a relationship between said
tracker and advertiser. KASHF uncovers several tracker advertiser relationships
which were previously unknown. The sustainability of KASHF takes a hit as
publishers migrate towards server side Header Bidding, eliminating the crucial
vantage point KASHF requires. Our work ATOM, builds upon KASHF, as an
ad-delivery / artifact-availability agnostic framework. {\em ATOM uses ad
creatives instead of bid values to infer relationships between trackers and
advertisers, awarding it immunity against changes in publisher or advertiser
practices.}
Common to both these prior approaches and our own is the concept of
`network tomography' \cite{Network-Tomo} as the fundamental measurement
approach. Essentially, these studies make inferences about the internals of the
ad ecosystem by predictably modifying it's input and monitoring the effects of
these modifications on the observable outputs. These pairs of inputs and
outputs are then used to make inferences about the unobservable internals of
the ecosystem. The idea has been used widely in the context of Internet
measurement. For example, Bu \etal \cite{Bu-SIGMETRIC02} leveraged the principles 
of network tomography to infer a networks internal link-level performance by 
analyzing end-to-end multicast measurements from a collection of trees. 
Similarly Castro \etal \cite{Castro-2004network} provided an overview of 
using network tomography to measure link and router level performances in 
large scale communication networks whereas, Lawrence \etal 
\cite{Lawrence-2006network} catalogs developments surrounding 
network tomography with a focus on active network tomography.

\section{Discussion}\label{sec:discussion}

\para{Evolution of the online advertising ecosystem.}
In recent years, users' privacy-awareness has significantly increased. This has
spurred many changes in the technologies and regulations surrounding user
data. 
Notably, regulators and developers of popular browsers have
sought to limit privacy-invasive behaviors by forcing transparency of data
handling practices and providing the user with more control over their data.
Unfortunately, this has resulted in the rapid development of technologies to
circumvent these protections. One such example is the development of
server-side cookie syncing solutions such as UnifiedID as a response to browser
limitations on tracking via third-party cookies.
Given the current push towards server-side solutions for advertising and
tracking, it is predictable that measurement of the online data ecosystem will
become more challenging.
ATOM aims to address this push towards server-side tracking and sharing
technologies by developing a framework for identifying data sharing
relationships even when they are not visible to the client. 
By leveraging the output of the advertising ecosystem rather than
a specific artifact within it, ATOM is expected to remain useful
despite the rapid churn in technology.

\para{Frameworks to improve regulatory enforcement.}
As user data and advertising continue to remain the primary monetization model
for the Internet, users can expect limited transparency and consent in to how
their data is being used. 
Recent regulatory efforts such as the GDPR, CCPA, and CDPA have sought to
remedy these harms. However, a major impediment to their effectiveness is the
inability to measure their violations and the limited resources available to
the bodies that are tasked with enforcing them.
It is crucial for researchers and governing bodies to invest in the development
of auditing frameworks to address these challenges.
ATOM contributes to this need by providing a general framework for
gathering statistically grounded evidence for potential violations of data
sharing disclosure regulations. 

\para{Limitations.} Fundamentally, ATOM is a best-effort attempt to develop
an artifact- and mechanism-independent method to identify data sharing
relationships in the online behavioral advertising ecosystem. It faces several limitations that impact its capabilities.

\parait{Scalability.} In our test deployment of ATOM, we face many challenges related to
scalability. Most notably, in order to account for the possibility of one
advertiser having sharing relationships with multiple trackers, we need to
consider all possible combinations of tracker blocking configurations ---
a challenge that resulted in the need to synchronize crawls for 1024 personas.
However, given the current oligopoly in the tracking ecosystem, we plan to
address this challenge with scale, by re-configuring ATOM to focus on a smaller
set of trackers and increasing our focus on ads from a larger number
of advertisers.

\parait{Completeness of ad corpus.} ATOM captures and identifies `ad images' only and depends on filter lists to identify these ads. We acknowledge that due to the absence of an exhaustive and 
programmatic ad identifying mechanism, our ad corpus is incomplete. We also plan 
to integrate other forms of media (video, gif, multilayered ads) to ATOM in the future.
However, our current results demonstrate that we can infer server side relationships by 
only using ad images with high significance.

\parait{Contract ads.} Since it is impossible to differentiate between 
programmatic and contract ads, we collect both. Programmatic ads are delivered 
via HB or RTB and facilitate personalized ads, whereas contract ads are fixed for a 
website irrespective of users. However, since contract ads are website specific, their 
presence in each persona would assign them low significance in our analysis.

\parait{Simplistic ad features.} We rely on very simple, yet seemingly effective features, extracted
from ad creatives by the Google Vision API. This decision was made based on the
manual validation from our pilot experiment concerning the quality of other
extracted features. It remains unclear if more advanced image processing tools
might improve the performance of ATOM.

\parait{Probabilistic results.} As described in \Cref{sec:relationships:methods:learning}, this research effort is complicated by the inherently probabilistic nature of online
advertising. This restricts our ability to make claims of definite
relationships between trackers and advertisers. Instead, we can only provide
statistically sound evidence of these relationships. We address this limitation
by configuring ATOM to conservatively infer errors. 

\parait{Lack of validation mechanisms.} Given the inability to validate the correctness of all our inferences, we limit our use of ATOM as a tool to inform stakeholders of {\em potential
violations} of disclosure regulations and motivate deeper investigations.

\para{Conclusions.} 
In this paper, we presented ATOM --- an artifact- and protocol-independent
mechanism for identifying data sharing relationships in the online advertising
ecosystem. 
ATOM is built on the insight that personalized ads themselves contain
information about an advertiser's knowledge of a user's activities. We
demonstrate the validity of this insight (\Cref{sec:interests}) and then
operationalize it (\Cref{sec:relationships}) to uncover data sharing
relationships, including those not visible to a user's browser and those not
discovered by any existing methods.
\section{Acknowledgement}
\label{sec:acknow}

We would like to thank the anonymous reviewers, as well as our shepherd, Tobias Urban, for their helpful suggestions. This research is supported by a Google Faculty Research Award.

\newpage
\bibliographystyle{ACM-Reference-Format}
\bibliography{ref}
\listoffixmes

\end{document}